\documentclass[conference]{IEEEtran}
\IEEEoverridecommandlockouts
\usepackage{cite}
\usepackage{amsmath,amssymb,amsfonts}
\usepackage{algorithmic}
\usepackage{orcidlink}
\usepackage{graphicx}
\usepackage{multirow}
\usepackage{textcomp}
\usepackage{url}
\usepackage{cite}
\usepackage{xcolor}
\def\BibTeX{{\rm B\kern-.05em{\sc i\kern-.025em b}\kern-.08em
    T\kern-.1667em\lower.7ex\hbox{E}\kern-.125emX}}
\begin{document}

\title{IoT-based Cost-Effective Fruit Quality Monitoring System using Electronic Nose}

\author{\IEEEauthorblockN{Anindya Bhattacharjee\orcidlink{0009-0006-6496-7896}, Nittya Ananda Biswas\orcidlink{0009-0009-8608-5714}, Khondakar Ashik Shahriar\orcidlink{0009-0008-2571-7485}, Kawsain Bin Salim}
\IEEEauthorblockA{\textit{Department of Electrical \& Electronic Engineering} \\
\textit{Bangladesh University of Engineering \& Technology}\\
Dhaka, 1205, Bangladesh \\
}
}
\maketitle

\begin{abstract}
Post-harvest losses due to subjective quality assessment cause significant damage to the economy and food safety, especially in countries like Bangladesh. To mitigate such damages, objective decision-making backed by scientific methods is necessary. An IoT-based, cost-effective quality monitoring system can provide a solution by going beyond subjective quality monitoring and decision-making practices. Here, we propose a low-power, cost-effective fruit quality monitoring system with an array of MQ gas sensors, which can be used as an electronic nose. We track the volatile gas emissions, specifically ethanol, methane, and ammonia, encompassing both ripening and decomposition for a set of bananas. Based on the gas concentration thresholds, we develop a mathematical model to accurately assess fruit quality. We also integrate this information into a dashboard for prompt decision-making and monitoring to make it useful to the farmers. This approach has the potential to reduce economic losses, enhance food safety, and provide scalable solutions for the supply chain.
\end{abstract}


\section{Introduction}

Agriculture, which contributes 11.66\% of the total GDP and employs 45.33\% of the workforce~\cite{stat_yearbook_bangladesh_2023}, is an important part of Bangladesh's economy. Traditional farming practices in rural Bangladesh are highly dependent on experience and generational knowledge, with limited adoption of modern technological advances. The dependence on subjective decision-making in post-harvest handling and crop management leads to large amounts of post-harvest losses, estimated at 20-30\% in developing economies like Bangladesh~\cite{food_losses_bangladesh_2021}. These losses pose a significant threat to food security and economic stability for rural farmers. Thus, smart and cost-effective post-harvest handling and quality monitoring beyond traditional practices are of great importance and are also in huge demand in developing countries.

Different types of quality assessment methods exist, of which traditional visual and manual assessment methods dominate the most in rural communities. Farmers check the size, shape, visual defects, firmness, and spoilage odors of fruits and estimate fruit quality based on experience. Although this method is cheap, it is very subjective and unreliable~\cite{ahmad2015postharvest}. Recent advances in microelectronics allow for small, handheld devices to non-destructively assess ripeness, defects, or internal quality by measuring light reflection or transmission. However, such fruit quality meters are very costly and require training for familiarization with the technology. Development of intelligent packaging, such as a film that changes color in response to spoilage gases, or a film that absorbs ethylene, can provide both assessment and preservation at low cost. However, these are typically done as part of small-scale interventions or pilot projects in middle-income countries~\cite{kalaitzis2016innovative}. Such technologies are hard to implement on a large scale and may not be the most cost-effective solution. Simple smartphone attachments or apps are also appearing as assessment aids, improving access to digital image-based analysis using machine learning and deep learning~\cite{knott2023facilitated, YUAN2024100656}. However, image-based techniques relying on deep learning are computationally intensive and power hungry. Besides image-based analysis, Zakaria et al. showed classification of mangoes' maturity and ripeness levels using the data of an electronic nose (E-nose) and an acoustic sensor through principal component analysis and linear discriminant analysis~\cite{Zakaria2012}. Mallick et al. show a graphene oxide sensor for ethylene gas detection from 40 to 120 ppm to detect the ripening of fruits such as guavas and bananas~\cite{Mallick2023}. Metal oxide semiconductor-based gas sensors have been used to predict fruit maturity for peaches in pre-harvest conditions, and machine learning methods like RNN have been implemented to determine the quality~\cite{voss2020nose}. But such high-quality sensors, implemented with machine learning for classification, lessen the affordability of the device. Implementation of MQ sensor-based E-nose to predict mango quality during the ripening stage is an affordable solution to fruit quality monitoring~\cite{rahman2024prediction}. The study shows a positive correlation between the soluble sugar content (SSC) and the hardness of mangoes and the readings of the MQ gas sensor during the ripening stage. Then it uses linear regression techniques to measure the hardness and SSC from the sensor data. Gonzales et al. used an array of MQ gas sensors to determine the ripeness of the Jackfruit and to classify it into three categories: unripe, ripe, and forcibly ripe, and implemented a machine learning model to predict the fruit quality with 96\% accuracy~\cite{gonzales2023gas}. In another study, an electronic nose composed of MQ-3, MQ-6, MQ-8, and MQ-135 was used to detect the various odors emitted by fruits, including apples, lemons, and bananas. An image processing method observing the shape and color was also used to monitor the fruits' conditions in parallel~\cite{suthagar2021determination}. 
\begin{figure}[b]
    \centering
    \includegraphics[width=0.8\linewidth]{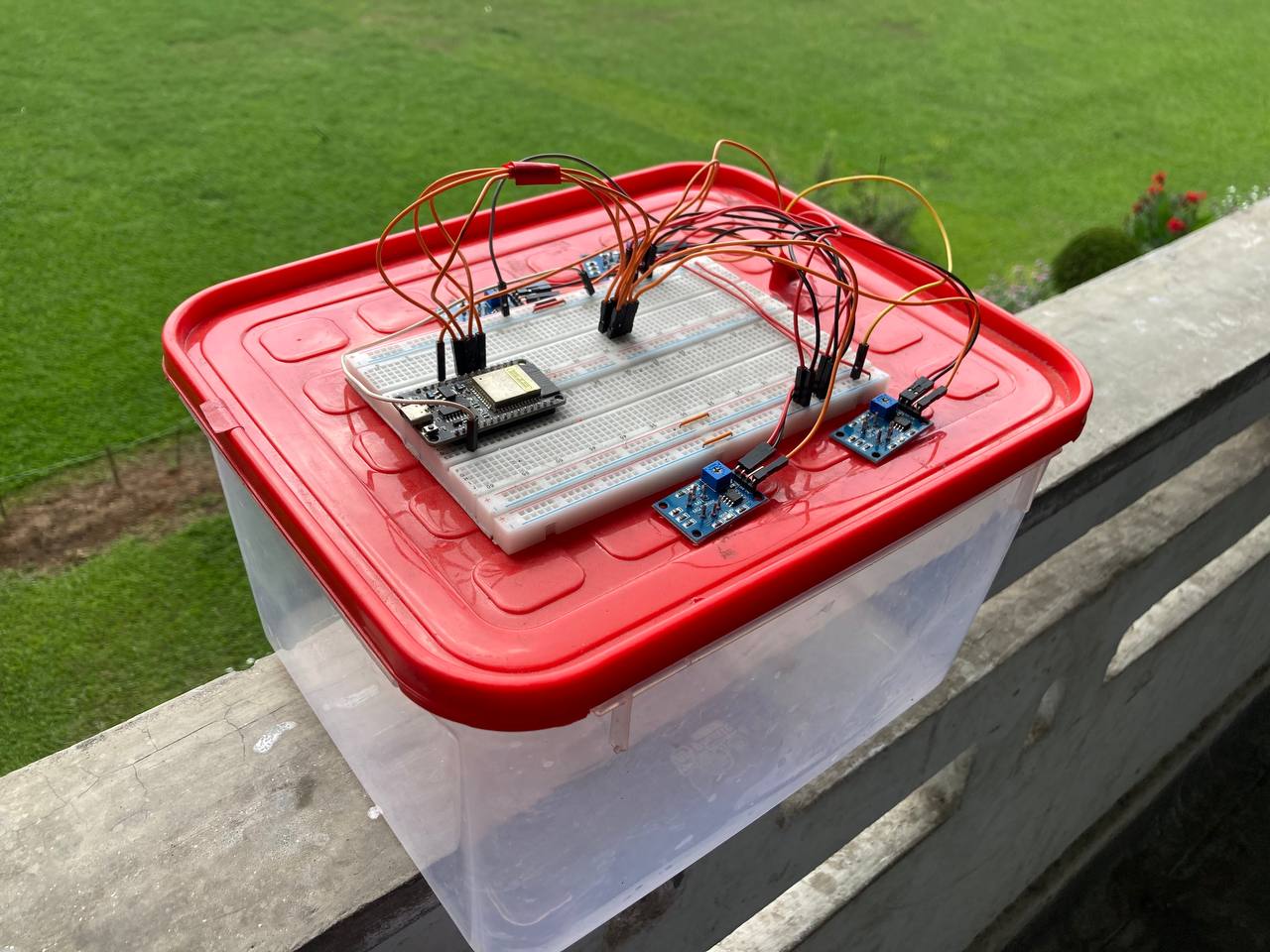}
    \caption{Sealed container setup for monitoring gas concentration}
    \label{setup}
\end{figure}

In our study, we observe the changes in ethanol, methane, and ammonia concentration in a sealed container using an array of MQ sensors (MQ-3, MQ-4, and MQ-135). We also monitor the temperature and humidity of the storage conditions. The contribution of our work is threefold. Firstly, we have built the model using cost-effective and readily available sensor arrays and equipment. Secondly, we have assigned a quality index based on gas concentration thresholds through mathematical modeling without the use of computationally intensive machine learning techniques. Lastly, to make the results more interpretable to rural farmers, we have built a dashboard that shows the real-time quality of fruits during storage conditions.

\section{Methodology}

\subsection{System Design: The E-nose}
The heart of our system is an ESP32 microcontroller with a built-in WIFI module, which connects all the MQ sensors (MQ-3, 4, and 135) and the temperature and humidity sensor. The sensor's core is a sensing element coated with tin dioxide ($SnO_2$) on an aluminum oxide ceramic substrate. The sensor uses a heater to heat the element to 200-300°C for gas detection. When target gases interact with the $SnO_2$ surface, they react with adsorbed oxygen, releasing electrons, and reducing resistance proportional to gas concentration. Note that each MQ sensor can detect multiple gases; that is, its resistance can vary with different chemical reactions. However, the sensitivity of each gas is not the same. Each gas sensor has a gas it is most sensitive to, which acts like its target gas. We refer to the datasheet of each sensor, which indicates that MQ-3, 4, and 135 sensors are the most sensitive to ethanol($C_2H_5OH$)~\cite{mq3b}, methane($CH_4$)~\cite{mq4}, 
and ammonia($NH_3$)~\cite{mq135}, respectively. Table~\ref{Gas Sensor} shows the detection range and the maximum power consumption for each of the gas sensors.

\begin{table}[htbp]
    \centering
    \caption{Gas Sensor Details}
    \label{Gas Sensor}
    \begin{tabular}{c c c c}
    \hline
    \textbf{Sensor} & \textbf{Target Gas} & \textbf{Detection Range} & \textbf{Power Consump.}  \\
    \hline
    MQ-3~\cite{mq3b} & Ethanol & 25-500 ppm & $\leq$ 900 mW\\
    MQ-4~\cite{mq4} & Methane & 300-10,000 ppm & $\leq$ 1 W \\

    MQ-135~\cite{mq135} & Ammonia & 10-1000 ppm & $\leq$ 950 mW\\
    
    \hline
    \end{tabular}
\end{table}

\begin{figure}[t]
    \centering
    \includegraphics[width=1\linewidth]{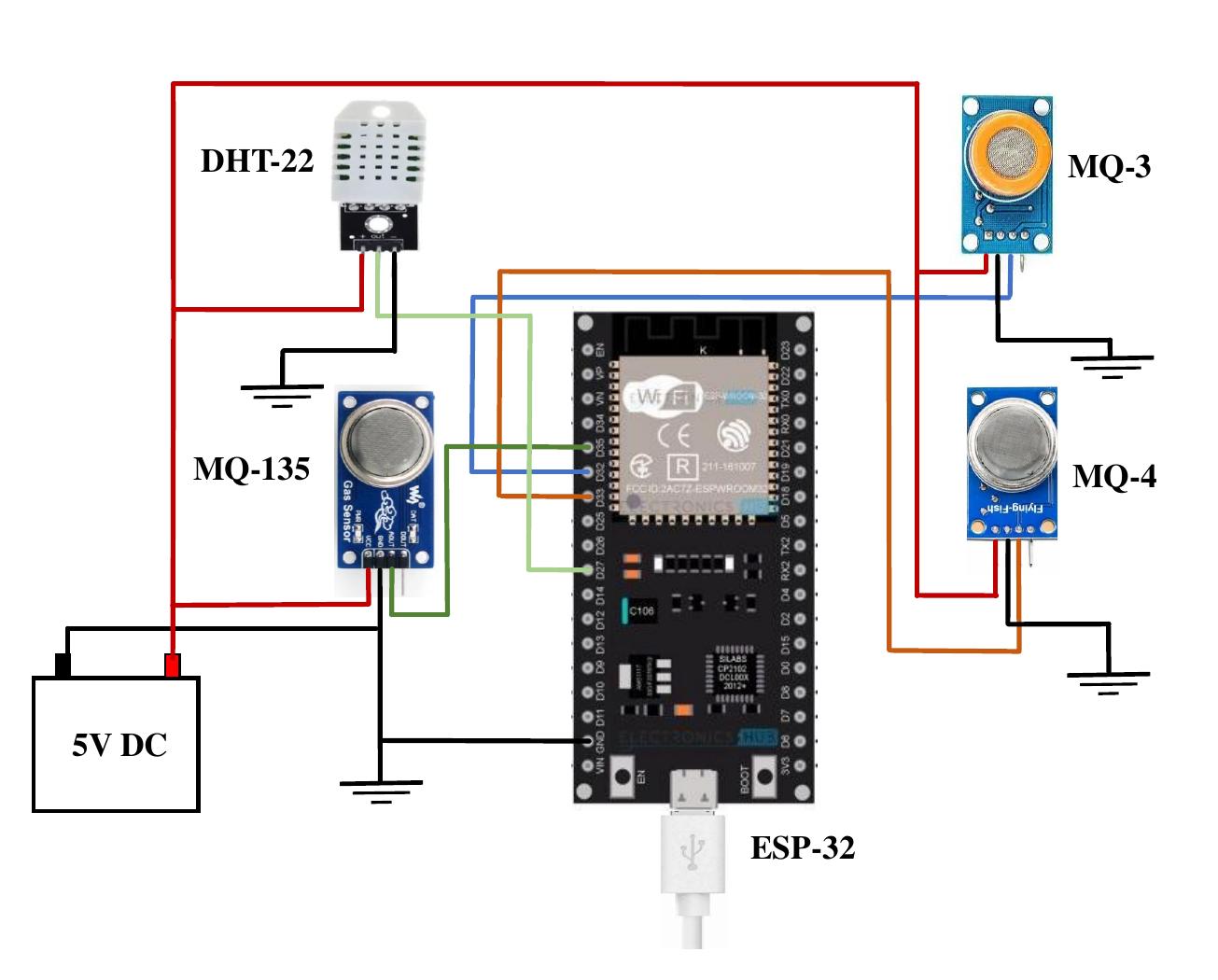}
    \caption{Circuit diagram of the sensor array of the E-nose}
    \label{Sensor ensemble}
\end{figure}

Each datasheet contains a sensitivity curve for the target gas that specifies the resistance ratio ($\frac{R_S}{R_o}$) vs. gas concentration curves, where $R_s$ is the sensor resistance at different concentrations and $R_o$ is the resistance in clean air~\cite{mq3b,mq4,mq135}. The sensor produces an output in voltage that changes proportionally as the resistance changes due to changing gas concentrations. A DHT-22 sensor is also incorporated in the system to monitor the temperature and humidity of the storage environment. The sensors are powered with a 5V DC supply and have a combined maximum power requirement of 4W. The setup of the sensor array connected to the ESP32 microcontroller is mounted on the top of the sealed container of dimension 28.5$\times$24$\times$17 cm$^3$, as shown in Figure~\ref{setup}, and the connection diagram of the sensors with the ESP32 microcontroller is shown in Figure~\ref{Sensor ensemble}. Based on local market prices in Bangladesh as of 2025, the whole setup costs approximately 2200 BDT or 18 USD.

From the output of the voltage values of the sensors, we can estimate the change in sensor resistance over time. By utilizing the curve of the resistance ratio vs gas concentration provided on the sensor datasheet, we identify the gas concentration in ppm~\cite{mq3b, mq4, mq135}. Note that we cannot state with absolute certainty that the resistance change in the gas sensor is exactly caused by the specified target gas, as sensor resistance can be varied to some degree by multiple gases. However, we state that the target gases detected by the sensors are expected from the fruit specimens during their ripening and decomposition stages~\cite{foods12213966}.

\subsection{Data Acquisition}
The study primarily focuses on bananas due to their high post-harvest losses (20.3\%)~\cite{food_losses_bangladesh_2021}, economic significance, and their distinct volatile gas emission profiles suitable for MQ sensor detection. Fresh bananas were brought from the market and weighed. Then, they were placed in a sealed container with the three MQ sensors and a DHT22 sensor inside. The data from each sensor was taken every minute and logged into a Google sheet for more than three days. 
During this period, the gas concentrations for the ripening and decomposition stages of bananas were logged and used for quality assessment. The weight of the bananas was 765g. Since the gas emitted by the fruits is likely to increase with the fruit quantity, the complete analysis of the gases was done using ppm/kg units. Though the assumption of the gas concentration rising linearly with weight may not be valid in all cases, it is a reasonable assumption for simplification. 

\subsection{Data Reliability}
Using low-cost sensors has its own challenges, as the readings can be faulty and noisy. To ensure the reliability of gas sensor readings, we conducted a comprehensive signal quality analysis. The evaluation included noise-related measures like signal-to-noise ratio (SNR), residual noise, and rolling standard deviation. We have fitted a polynomial curve as the baseline signal, and then treated deviations as noise when calculating SNR. For rolling standard deviation, we used a window size of 120 samples. We also calculated temporal characteristics like lag-1 autocorrelation. These metrics quantify sensor precision, predictability, and robustness against noise. In this context, we propose that a high signal-to-noise ratio indicates that the measured ripening trends dominate over random fluctuations. Also, high autocorrelation reflects temporal consistency in the sensor response, which we evaluate as a demonstration of the practical reliability of the proposed E-nose system.

\subsection{Quality Assessment}

The quality assessment depends on the concentration thresholds for ripening and decomposition. We assign a normalized quality index for each of the gases and combine the indices using a weighted average. For the quality index, we introduce the formula as,
\begin{equation}
    Q_{gas} = max(0, 1 - \frac{x^a}{b^a})
    \label{Gas quality}
\end{equation}
where $a$ and $b$ values are determined from the gas concentration values at ripe and rotten stages. The value for $b$ is equal to the concentration threshold for decomposition. The value of $a$ depends on the ripening threshold, so that the index stays at 0.98 at the threshold. The $max()$ function ensures that the quality index drops to zero beyond the decomposition threshold. 
\begin{figure}[t]
    \centering
    \includegraphics[width=0.9\linewidth]{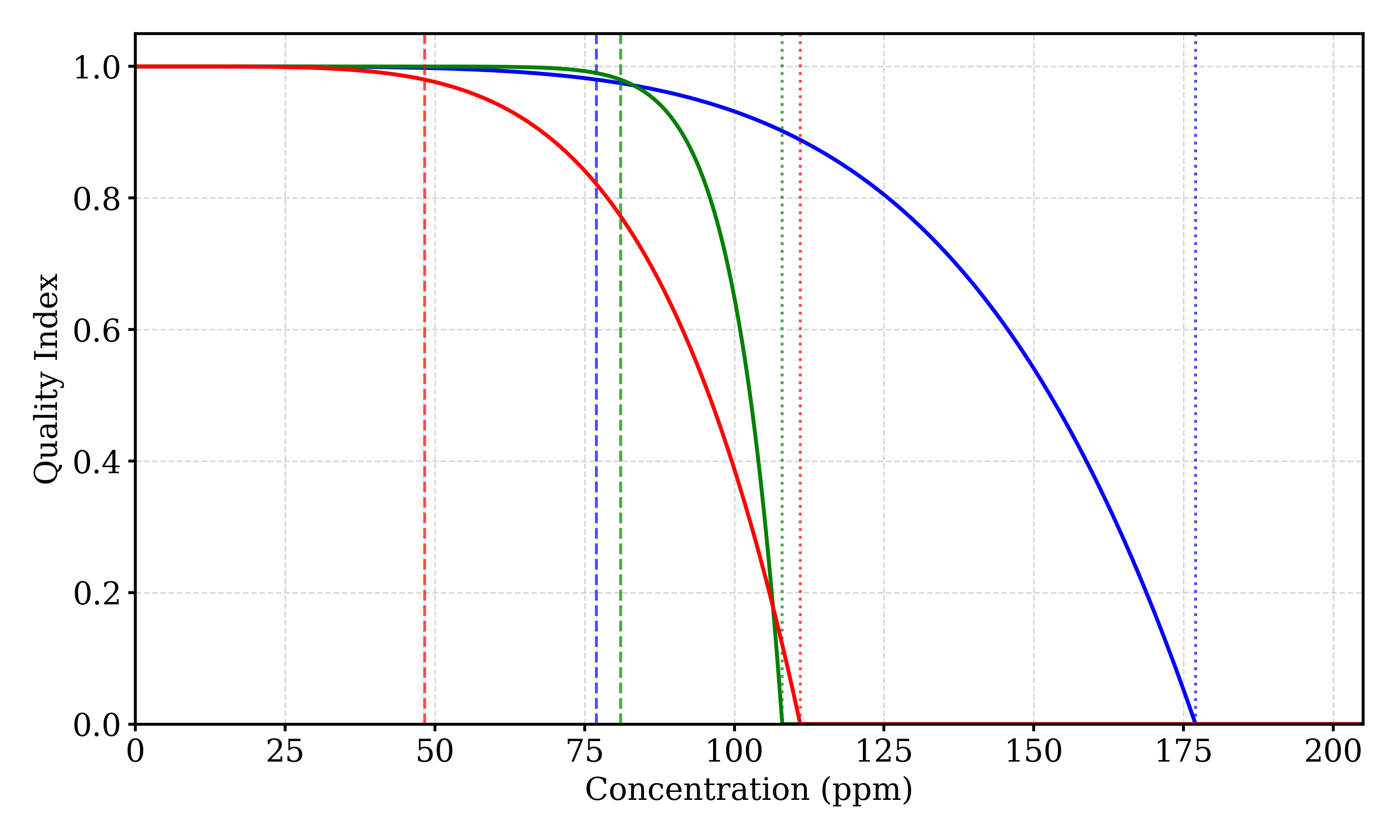}
    \caption{Different quality index degradation of gases based on a,b constant values}
    \label{modeling quality}
\end{figure}
In Figure~\ref{modeling quality}, we can see three types of quality index curves with dashed vertical lines indicating the threshold for the ripe stage and dotted vertical lines indicating the threshold for the decomposed stage. The blue curve indicates fruits that have a high ripening and high rotting time, such as mangoes, avocados~\cite{hofman2001ripening}. The green curve represents fruits that have a high ripening time but decompose very fast after being ripe. Berries like strawberries and raspberries follow this trend~\cite{umass_postharvest_storage}. Finally, the red curve has both faster ripening and decomposition, which can be seen in bananas~\cite{ruwali2022effect}. This breakdown of fruit types is a simplification of a complex biological process, but it's a useful model for understanding their behavior. Depending on the fruit, we can adjust the constants to properly model their condition over time.

In addition to gas concentrations, storage conditions such as temperature and humidity in the environment also contribute to the quality of the fruit. Bananas have a very short post-harvest shelf life and are also very sensitive to cold temperatures less than $13^o$C. At ambient conditions, their shelf life is four to ten days when mature green and two to four days when ripe~\cite{mejia2005handling}. So, for bananas, the ideal storage temperature is $14^o -16^o$C and 90-95\% relative humidity. To add the effect of storage conditions, we also add a quality index for temperature and relative humidity.

\begin{equation}
    Q_{temp} = \begin{cases}
        ~max(1 - \frac{T_{min}-T}{S_{T_1}},0);~T<T_{min} \\
        ~1;~~~~~~~~~~~~~~~~~T_{min}\leq T \leq T_{max} \\
        ~max(1 - \frac{T-T_{max}}{S_{T_2}},0);~T>T_{max} \\
    \end{cases}
\end{equation}
\begin{equation}
    Q_{RH} = \begin{cases}
        ~max(1 - \frac{H_{min}-H}{S_{H_1}},0);~H<H_{min} \\
        ~1;~~~~~~~~~~~~~~H_{min}\leq H \leq H_{max} \\
        ~max(1 - \frac{H-H_{max}}{S_{H_2}},0);~H>H_{max} \\
    \end{cases}
\end{equation}

Here, $S_T$ and $S_H$ are quality tolerance factors for out-of-range temperature and relative humidity, and the subscripts $1$ and $2$ indicate lower and higher conditions than the threshold.

\begin{table}[htbp]
    \centering
    \caption{Tolerance Values for Quality Indices of different environmental conditions}
    \label{sensitivity values}
    \begin{tabular}{c c c c}
    \hline
       \textbf{Fruit} & \textbf{Factors} & \textbf{$S_1$} & \textbf{$S_2$}\\
       \hline
        \multirow{2}{*}{Banana} & Temperature & 2 & 9 \\
        & Humidity & 10 & 5 \\ 
       \hline
    \end{tabular}
\end{table}
In Table~\ref{sensitivity values}, we can see that the tolerance of bananas for low temperatures is quite low compared to the tolerance for high temperatures. The high relative humidity tolerance has a maximum value of 5\% since the ideal maximum relative humidity itself is 95\%. Low relative humidity tolerance is larger, allowing relative humidity to go as low as 80\% from the ideal minimum 90\%.
Now, the final quality is a weighted average of all the quality indices.

\begin{equation}
    Q_{total} = \sum_{i=1}^{5}{w_iQ_i} ~;~~ \sum_{i=1}^{5}w_i = 1
\end{equation}

This average quality shows a combined impact of the factors that are a good representation of the fruits' changes over time. We note that the factor changing more rapidly over time for gases is a defining factor of quality and thus must be assigned a higher weight compared to the factor that changes slowly. 

\subsection{Data Representation}
To effectively present the data to the users of this prototype, especially to the rural farmers, we have made a dashboard that shows the weight stored in the container, the number of active sensors, the current sensor readings, as well as their past trends~\cite{Bhattacharjee_2024}. The instantaneous quality score with five different quality categories, Excellent, Good, Moderate, Bad, Rotten, is also directly shown for ease of understanding. Since the project is targeted at rural farmers in Bangladesh, there are both Bengali and English versions of the dashboard.

\section{Results}

During the data acquisition for the bananas, the storage temperature was $32^o$C and the relative humidity was 97\%. Since the storage temperature was far from ideal, we have observed a low shelf life for our bananas, compared to the typical shelf life stored in low temperatures. 
\begin{figure}[t]
    \centering
    \includegraphics[width=0.9\linewidth]{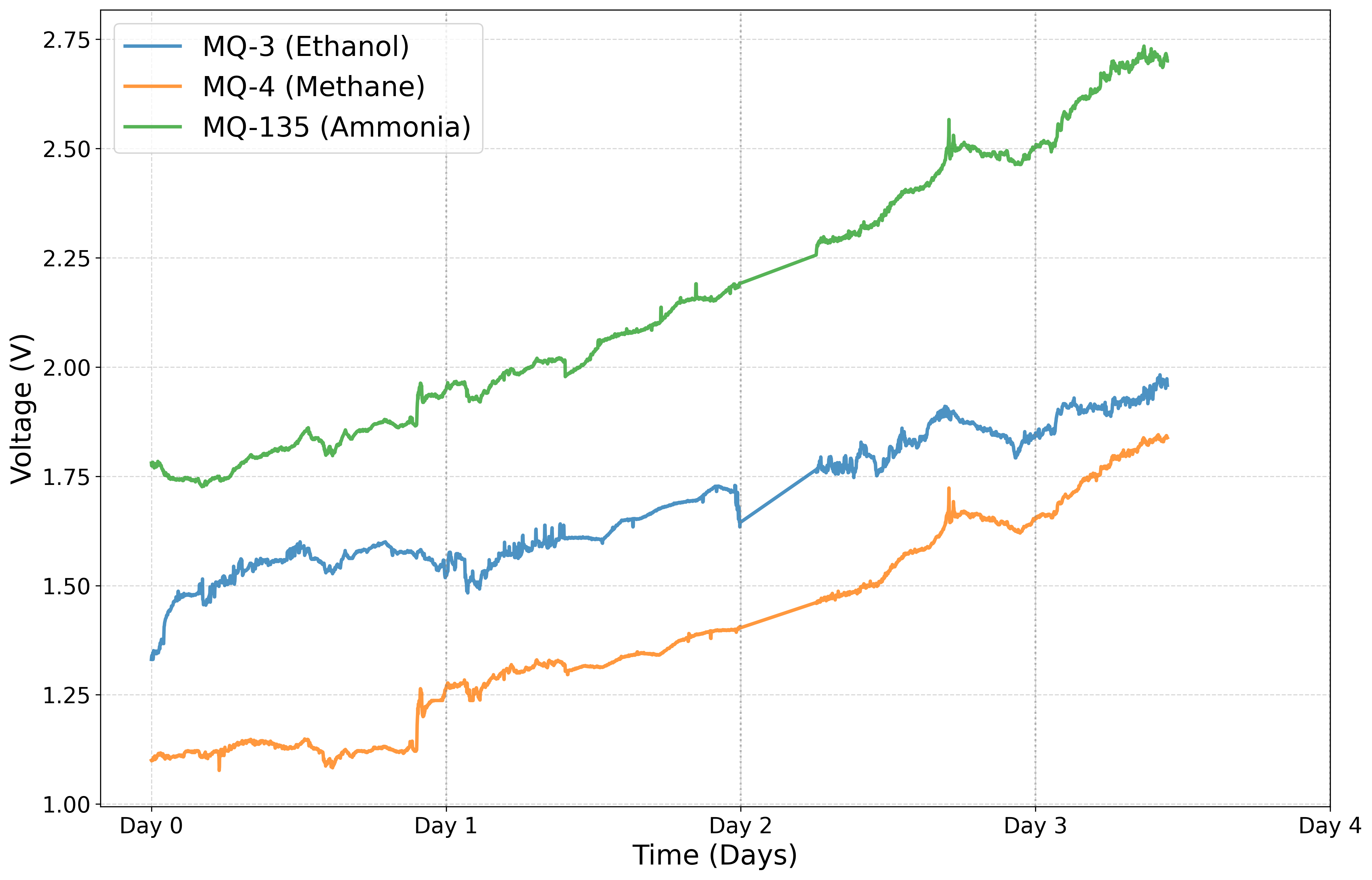}
    \caption{Voltage output by the sensors over time}
    \label{voltage reading}
\end{figure}

\begin{figure}[b]
    \centering
    \includegraphics[width=0.9\linewidth]{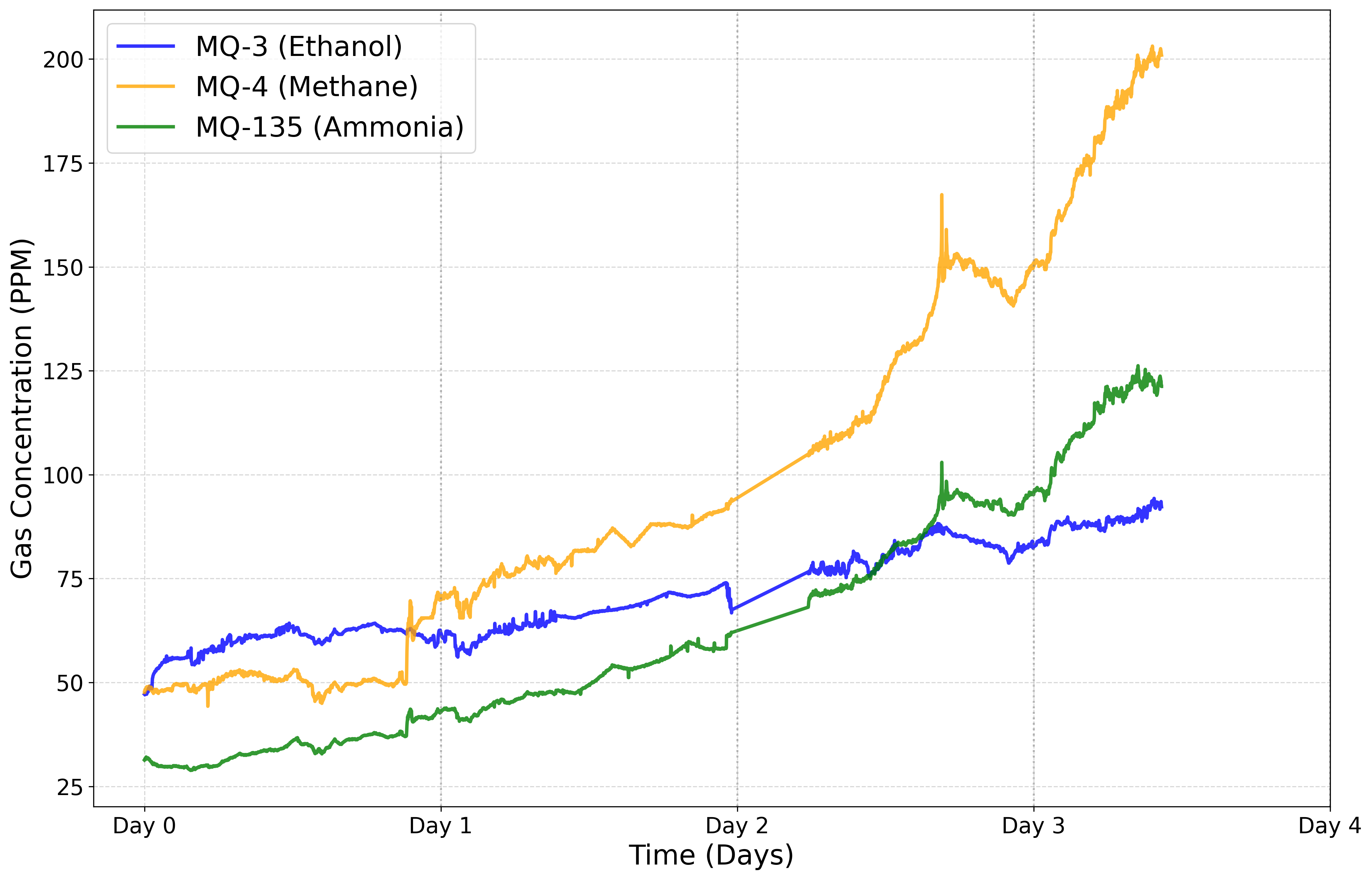}
    \caption{Change in gas concentration with time}
    \label{ppm reading}
\end{figure}

From Figure~\ref{voltage reading}, we can see that the voltage output by the gas sensors rises over time. Converting the designated voltage level to ppm using the sensitivity curve in the datasheet~\cite{mq3b,mq4,mq135}, we can observe the changes in the gas concentration. 

In Figure~\ref{ppm reading}, we can see that all of the gases have increased concentration over time, with methane and ammonia having a high growth. On day 1, the bananas began to ripen, which was reflected in the gas concentration trends. Also, between days 2 and 3, we see a rise in the concentration of both methane and ammonia, indicating the start of decomposition.

 
\begin{table}[htbp]
    \centering
    \caption{Concentration thresholds for banana at different stages}
    \label{concentration threshold}
    \begin{tabular}{c c c}
    \hline
       \multirow{2}{*}{\textbf{Gases}} & \textbf{Ripe stage} & \textbf{Decomposed stage}\\
       & (ppm/kg) & (ppm/kg)\\
       \hline
        Methane & 92 & 177 \\
       Ethanol & 81 & 108 \\
       Ammonia & 48 & 111 \\
       \hline
    \end{tabular}
\end{table}

The sealed container was transparent so that we could observe the color changes in the banana over time. For the ripe stage, we looked for a solid yellow color. Brown spots started to occur on the overripe banana, and when there were dark brown patches, the bananas were considered at the beginning of decomposition. When certain areas were black and created a rotten smell, the bananas were considered decomposed. From such visual information, we determined the thresholds for the ripe state and decomposed state, which are shown in Table~\ref{concentration threshold}.  

Based on these thresholds, we fit the quality index to equation~\ref{Gas quality}. Each gas has different thresholds, which dictate the constant values as shown in Table~\ref{constant values}. The threshold value should also vary for different fruits.

\begin{figure}[t]
    \centering
    \includegraphics[width=0.9\linewidth]{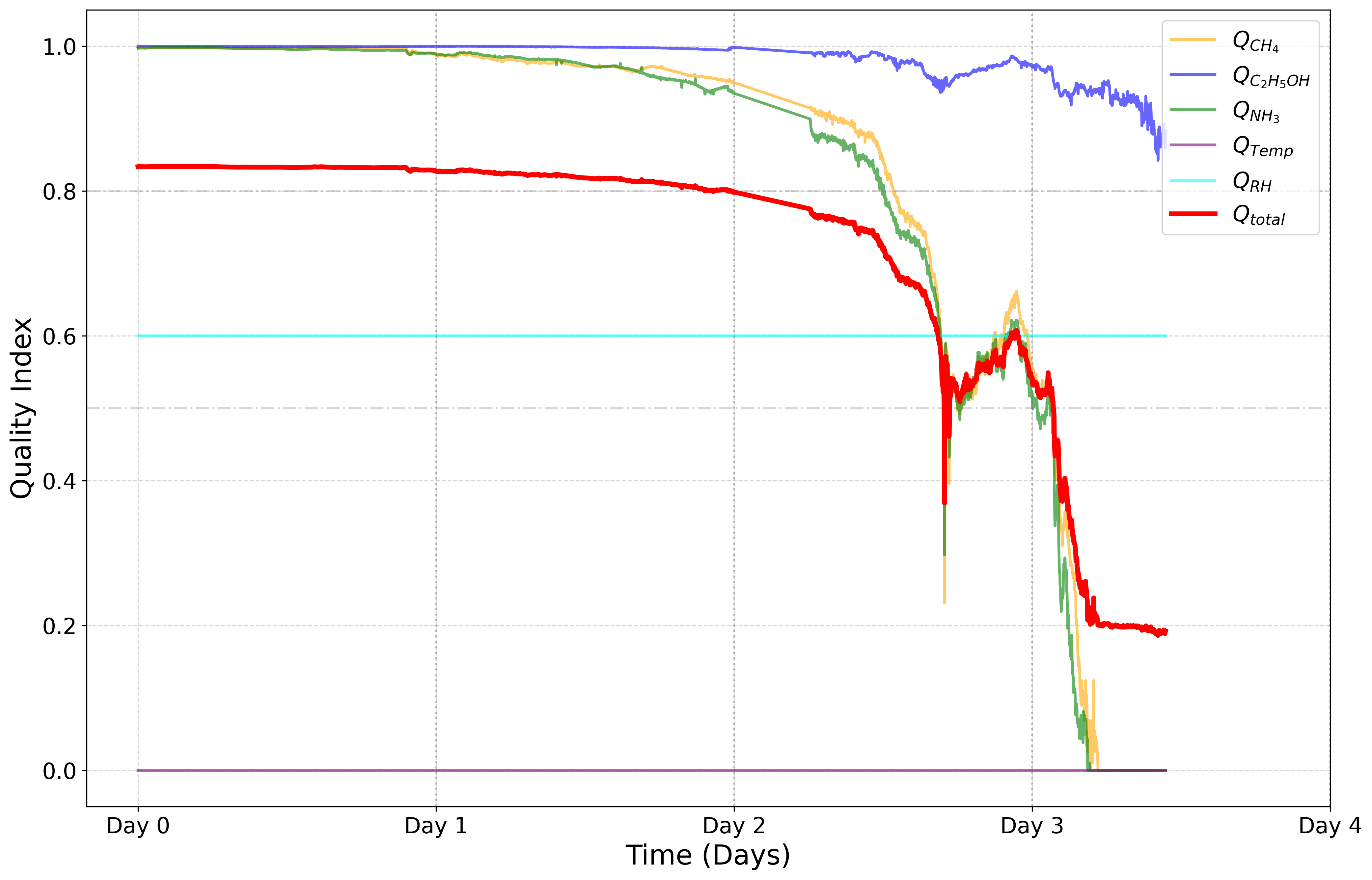}
    \caption{Individual and combined quality index for banana over time}
    \label{Quality index banana}
\end{figure}

\begin{table}[htbp]
    \centering
    \caption{Constant Values for Quality Indices of different gases}
    \label{constant values}
    \begin{tabular}{c c c c}
    \hline
       \textbf{Fruit}  & \textbf{Gases} & \textbf{a} & \textbf{b}\\
       \hline

       \multirow{3}{*}{Banana} & Methane & 4.7 & 177 \\
         & Ethanol & 13.6 & 108 \\
         & Ammonia & 4.7  & 111 \\
       \hline
    \end{tabular}
\end{table}

\begin{table*}[h]
\centering
\caption{Signal Quality Metrics of MQ-Series Gas Sensors}
\begin{tabular}{c c c c c c c}
\hline
\multirow{2}{*}{\textbf{Sensor}} & \multirow{2}{*}{\textbf{SNR} ($\frac{V}{V}$)} & \textbf{Residual Noise} & \textbf{Mean Rolling} & \multirow{2}{*}{\textbf{Autocorrelation}}\\
& & \textbf{Level} (V) & \textbf{Standard Deviation} & \\
\hline
MQ-3 (Ethanol) & 19.64 & 0.0031 & 0.0108 & 0.999 \\
MQ-4 (Methane) & 24.73 & 0.0311 & 0.0078 & 1\\
MQ-135 (Ammonia) & 44.24 & 0.0228 & 0.0097 & 1 \\
\hline
\end{tabular}
\label{tab:sensor_quality}
\end{table*}
After that, we estimate the reduction in quality for bananas based on the quality index for individual gases and environmental factors. In Figure~\ref{Quality index banana}, we can see the quality index dropping between day 2 and 3, which is when the banana started to decompose from its ripe condition. For methane, ammonia, and ethanol, the assigned weight for combined quality is 0.3, 0.325, and 0.15, respectively, as the former two undergo significant changes compared to the latter. For quality indices of environmental conditions like temperature and humidity, we assign a weight value of 0.125 and 0.1, respectively. This is because they do not directly represent the current quality state but act as an indicator of how well the preservation is slowing down the decomposition process.

\begin{figure}
    \centering
    \includegraphics[width=0.9\linewidth]{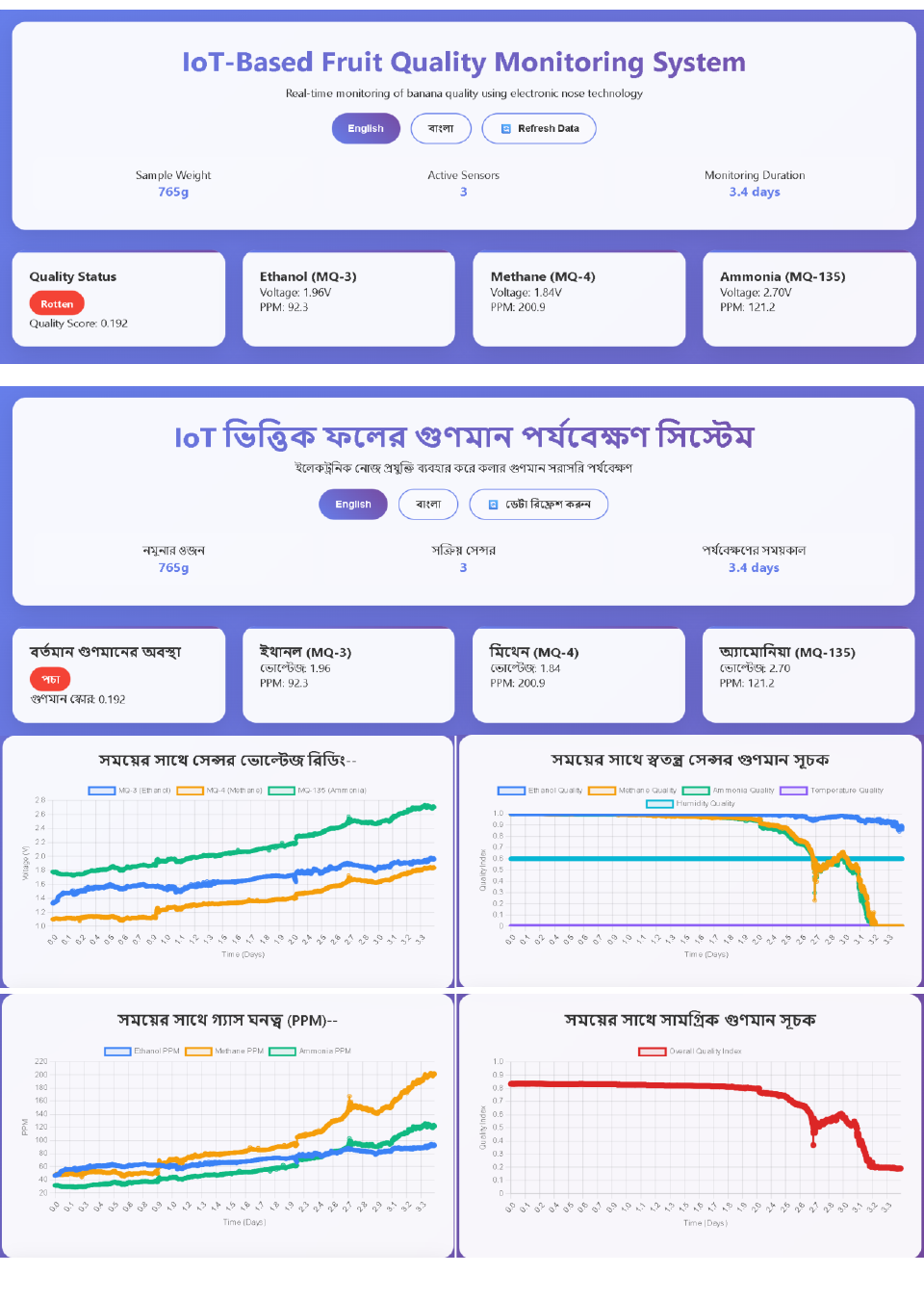}
    \caption{Dashboard representing the sensor data}
    \label{Dashboard}
\end{figure}
For ensuring sensor reliability, we analyzed the voltage readings of the sensor over time with some signal processing techniques. As shown in Table~\ref{tab:sensor_quality}, the MQ-135 sensor has the highest signal-to-noise ratio (SNR) and perfect autocorrelation, indicating superior temporal stability with minimal noise, making it the most reliable sensor in the group.  The MQ-4 sensor has a moderate SNR and nearly perfect autocorrelation, indicating high stability and precision.  In contrast, MQ-3 has the lowest SNR; despite being reasonably predictable, it may be more susceptible to noise changes. This is also another reason for assigning a higher weight in the combined quality assessment to the MQ-4 and MQ-135 sensors.
Finally, in Figure~\ref{Dashboard}, an image of our dashboard is provided, which helps visualize the data in both Bengali and English. The page can be updated through the \textit{refresh data} option, and the trends of gas concentrations can be seen as well, with a clickable value display for each reading in the graphs. 

\section{Discussion}
The major consideration in our development of the E-nose is to make it cost-effective and easily implementable, which is achieved by creating a prototype for only 18 USD. Compared to previous MQ sensor-based studies ~\cite{rahman2024prediction}, ~\cite{gonzales2023gas}, ~\cite{suthagar2021determination}, our work distinguishes itself through the combination of mathematical threshold modeling instead of regression or machine learning techniques, integration with a bilingual real-time dashboard for direct farmer accessibility, and explicit focus on cost-effectiveness for rural deployment. This approach trades classification accuracy for interpretability, reduced computational requirements, and immediate actionable insights. However, to maintain cost-effectiveness, we used MQ sensors. These are sensitive to environmental conditions such as sudden input voltage changes and mechanical vibrations, and they also require a long settling time before producing stable readings. The ppm values for gas concentration thresholds noted here depend on the design of the container. The gas concentrations can vary on the basis of their position of sensors in the setup and can be modified by leakages near sensor holes. Using advanced and more reliable sensors in a highly controlled environment can produce more accurate results. However, to ensure reliability, we observed the noise characteristics of the sensor data and found the SNR to be high for all the sensors. An improvement to this work could be achieved by replacing the current reliance on visual cues with more objective methods, such as chemical, spectroscopic, or other advanced analytical techniques, to precisely determine the ripe and decomposed states. Besides, introducing a weight loss factor over time and also scaling the IoT to the supply chain can be thought of as room for improvement. Experimentation on different types of fruits can also increase the generalizability of the study. In addition, the effect of different environmental conditions (temperature and humidity) on the senescence of the fruits should also be studied. Despite these issues, with proper design, implementation, and funding, this study can serve as a prototype for the development of a cost-effective electronic nose suitable for the quality monitoring of different fruits and crops upon proper calibration.

\section{Conclusion}
An E-nose is an innovative instrument utilized in the food industry for the swift, cost-effective, and non-invasive evaluation of quality. Such sensor arrays enable the recording of intricate odor profiles, which are challenging to accomplish with conventional techniques. Though sensor sensitivity highly impacts the degree of prediction accuracy of the E-nose device, it is one of the cheapest and most reliable ways of quality monitoring. Calibration of the system to other fruits, and integration with the necessary relevant sensors can make the system more versatile and universal. This type of cost-effective, non-invasive fruit quality monitoring system can have a good potential for many different applications. Such systems can be deployed in large storage facilities, embedded in refrigerator fruit compartments, or integrated into transport units. Our study serves as a prototype for these possible applications, which can reduce the losses in different stages of the supply chain, from the hands of farmers to the consumers.

\section*{Acknowledgment}
This study was conducted as part of the Microprocessor and Embedded Systems Laboratory project. The authors express their sincere gratitude to the Department of Electrical and Electronic Engineering of the Bangladesh University of Engineering and Technology for providing research support.

\bibliographystyle{ieeetr}
\bibliography{reference}

\end{document}